\documentclass[12pt]{iopart}
\usepackage{graphicx,color,iopams,setstack}

\newcommand{\ket}[1] {\vert{#1}\rangle}
\usepackage{setspace}
\begin{document}
\title[Simulation of gauge transformations on systems of ultracold atoms]{Simulation of gauge transformations on systems of ultracold atoms}
\author{O Boada, A Celi,
J I Latorre, V Pic\'o.}
\address{ Dept. d'Estructura i Constituents de la Mat\`eria,
Universitat de Barcelona, 647 Diagonal, 08028 Barcelona, Spain}

\begin{abstract}
We show that gauge transformations can be simulated on systems of ultracold atoms. We discuss observables that are invariant under these gauge transformations and compute them using a tensor network ansatz that
escapes the phase problem. We determine that the Mott-insulator-to-superfluid critical point is monotonically shifted as the induced magnetic flux increases. This result is stable against the inclusion of a small amount of entanglement in the variational ansatz.
\end{abstract}
\maketitle
\section{Introduction}

Ultracold gases on optical lattices provide a powerful instrument for creating
quantum devices that can simulate, in a controlled manner, a variety of condensed
matter systems. The emerging possibility of acting on the system with carefully
tuned electromagnetic fields invites the design of a new generation of
experiments \cite{LSADSS07}.  It is in principle possible to
implement synthetic Abelian \cite{JZ03,M04,SDL05} and non-Abelian \cite{OBSZL05} 
gauge fields coupled to \textit{neutral} ultracold atoms on optical lattices. For the off-lattice implementations see \cite{JO04,RJOF05}; for the rotation-based approach see \cite{BDZ08}. Recently, synthetic abelian fields for neutral atoms were successfully created \cite{YCPPPS09,YCJPS09,LCJPP10}. For a recent review on this topic see \cite{DGJO10}.

The main aim of this paper is twofold. First, we will show how the effective gauge field created for systems of neutral atoms propagating on an optical lattice can be transformed into other gauges by simply modifying the geometry of the experiment. Furthermore, we will show that it will be possible to test the invariance of the Hamiltonian, as proven by Lieb \cite{LiebLoss}, under these gauge transformations by setting up the experiment in different gauges and checking that certain measurements, in particular the location of the Mott-Insulator-to-Superfluid phase transition, yield the same result. More precisely, we will show that it is possible to simulate non-trivial pure gauge vector potentials and to verify that the physical effects they have are the same as those of a vector potential that is zero everywhere. 

In recent years there have been many proposals to build quantum simulators. Analogue simulators ---where the dynamics of a real physical system are approximated by another one in a lab--- have been proposed. Quantum circuits whose outputs are important ground states have been put forth \cite{LPR2009}. Also, devices that reproduce \textit{exactly} the dynamics of a real-world system have been presented \cite{LCV2008}. In this article, for the first time we propose the simulation of a gauge symmetry or, to be more precise, of a gauge transformation.

The second aim of this paper is to study the phase diagram of the Bose-Hubbard (BH) Hamiltonian with complex, position-dependent hopping amplitudes beyond mean field theory using a tensor network ansatz. This model is considered to be hard for Monte Carlo simulations due to the phase problem, a generalization of the fermionic sign problem, and is of great relevance for ultracold atomic systems on optical lattices. Although there is an extensive body of literature devoted to its study with mean field techniques, see Ref. \cite{fazio} and references therein, it has not been approached using methods that go beyond mean field for systems sizes that are beyond the reach of exact diagonalization. Using a projected entangled-pair state (PEPS) ansatz with bond dimension $\chi=1,2$ we compute the ground state and determine the quantum critical point of the Mott-insulator-to-superfluid phase transition \cite{sachdev} as a function of the magnetic flux going through the system. We find that the result for $\chi=2$, which allows for some degree of entanglement in the ground state, is not qualitatively different than that found with $\chi=1$, i.e. with a mean field ansatz.

The paper is organized in the following way. We will begin by reviewing the proposal by Jaksch and Zoller \cite{JZ03} to create an artificial magnetic field for neutral atoms on an optical lattice. In the next section we will discuss how the experimental setup can be modified in a simple way in order to transform the vector potential into a collection of other gauges. The last two sections will be devoted to introducing the relevant observable and to its computation using tensor network states.

\section{Synthetic gauge fields on an optical lattice}

Our starting point will be the Bose-Hubbard (BH) Hamiltonian, which describes bosonic atoms loaded on an optical lattice  \cite{JBCGZ98,SCMI02}. By increasing the laser intensity in one of the directions of the lattice it is possible to create an array of uncoupled two dimensional lattices, each of which is governed by a 2d BH Hamiltonian,
\begin{equation}
\eqalign{H=-\sum_{m,n=-\infty}^{\infty}\left(J_{x}a^{\dagger}_{m+1,n} a_{m,n}
+J_{y}a^{\dagger}_{m,n+1}a_{m,n} + H.c\right).+\cr
+\frac{U}{2}\sum_{m,n=-\infty}^{\infty}N_{m,n}(N_{m,n}-1) - \sum_{m,n=-\infty}^{\infty}\mu_{m,n}N_{m,n} \, ,}
\label{hubbard}
\end{equation}
where the bosonic operators $a^{\dagger}_{m,n}$ and $a_{m,n}$ create and destroy respectively an atom at a lattice
site $\bi{x}_{m,n}=a(m,n)$, with $m,n$ integers and $a$ being the lattice period. The
constant $J_x$ ($J_y)$ is the site-to-site tunneling energy in the $x$ ($y$)
direction. The parameter $U$ is the pair interaction energy at each site,
$\mu_{m,n}$ is the local chemical potential, and $N_{m,n}$ is the local
occupation. 


The next step is to make the lattice state-dependent. That is, let us suppose the atoms can be in two hyperfine states  $\ket{g}$ and $\ket{e}$, and that the lifetime of the excited state is such that spontaneous decays from $\ket{e}$ to $\ket{g}$ are negligible. Then, if the two states are trapped in different rows of the lattice, and we increase the laser intensity in the direction which is orthogonal to these rows we will obtain an array of uncoupled one dimensional lattices that trap atoms of each state in every other row.

Now, let us couple the states $\ket{g}$ and $\ket{e}$ by shining two Raman lasers on the system with wave vectors $\bi{k}_g$ and $\bi{k}_e$. It follows that an atom on the lattice will undergo Rabi oscillations between the two levels with a (crucially) position-dependent Rabi frequency
\begin{equation}
\Omega\left(\bf{x}\right)=\Omega_0 e^{i(\bi{k}_e - \bi{k}_g)·\bi{x}}\, ,
\end{equation}
where $\Omega_0$ is a constant depending on the Raman laser intensity and the detuning. When an atom undergoes a transition from $\ket{g}$ and $\ket{e}$ and vice-versa it finds itself in a maximum of the optical potential, and tends to fall towards a neighboring row of the lattice at a rate given by

\begin{equation}\label{raman_J}
J_R = \frac{1}{2}\int \omega^{*}\left(\bi{x} - \bi{x}_{m,n+1}\right)\Omega\left(\bi{x}\right)\omega\left(\bi{x} - \bi{x}_{m,n}\right)\,,
\end{equation}
where $\omega\left(\bi{x} - \bi{x}_{m,n}\right)$ is a Wannier function centered at site $\left(m,n\right)$. Setting $\bf{q}\equiv\bi{k}_e - \bf{k}_g$ parallel to the $m$ direction we obtain $J_R=J_0$exp$(2\pi i\alpha m)$, as in \cite{JZ03}, with $\alpha=\vert \bi{q}\vert a /(2\pi)$. 

However, allowing for $\bf{q}$ to point in any direction of the plane, that is, setting $\bi{q} = \vert \bi{q} \vert \left(\cos\theta, \sin\theta \right)$, the induced tunnelling rate in the $n$ direction becomes

\begin{equation}
\eqalign{
J_R = \frac{\Omega_0}{2} e^{iA_y(m,n)}&\int  dx\, \vert \omega_x\left(x\right) \vert^2 \cos\left(2\alpha\cos\theta\, x\right)\times  \cr
 &\times\int  dy\, \omega_y^{*}\left(y\right)e^{2\pi i \alpha\sin\theta }\omega_y(y-\frac{\pi}{2})\,,}
 \label{jram}
\end{equation}
where $A_y$ is the $y$-th component of the vector potential,
\begin{equation}\label{vector_potential}
A_y\left(m,n\right)=2\pi\alpha \left(\cos\theta\, m + \sin\theta\, n/2\right)\,, 
\end{equation}
plus an irrelevant constant. As we show in the appendix A, the integrals in $x$ and $y$ depend very weakly on $\theta$ and can thus be set to a constant. The final result is 
\begin{equation}
J_R=J_0 e^{2\pi i \alpha \left(\cos\theta\, m + \sin\theta\, n/2\right)}\, .
\end{equation}
Note that for $\sin\theta=0$ the model reduces to the Hofstadter Hamiltonian \cite{H76}.
The BH Hamiltonian for the system, because $\Omega_0$ can be tuned in order to set $J_x=J_0$, now reads

\begin{equation}
\eqalign{
H=-J\sum_{m,n=-\infty}^{\infty}\left(a^{\dagger}_{m+1,n} a_{m,n}+ e^{iA(m,n)} a^{\dagger}_{m,n+1}a_{m,n} + H.c.\right)+\cr
+\frac{U}{2}\sum_{m,n=-\infty}^{\infty}N_{m,n}(N_{m,n}-1) - \sum_{m,n=-\infty}^{\infty}\mu_{m,n}N_{m,n} \,.}
\label{hubbard2}
\end{equation}

\begin{figure}
\begin{center}
\scalebox{0.7}{\includegraphics{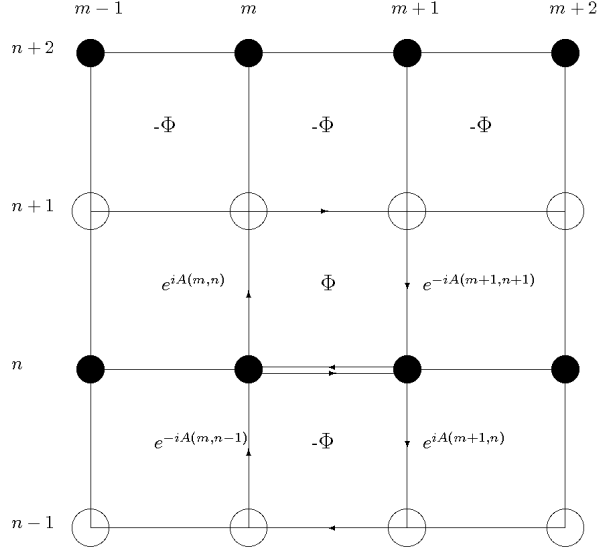}}
\end{center}
\caption{Circulation around a plaquette and phases picked up at each transition between lattice sites. The plaquettes in rows adjacent to the one in the diagram get opposite phases as in those cases the transitions between the two levels of the atoms take place in the other direction ( first form $\ket{e}$ to $\ket{g}$ as opposed to $\ket{g}$ to $\ket{e}$ ). This configuration gives rise to a staggered magnetic field. Filled circles trap atoms in state $\ket{g}$ and empty circles trap those in state $\ket{e}$.}
\label{plaquette}
\end{figure}

Hence, by coupling the two hyperfine states with Raman lasers and thus inducing transitions between them with a position dependent Rabi frequency, an atom going around a plaquette of the lattice will acquire a non-vanishing phase in the same way charged particles acquire an Aharanov-Bohm phase in a magnetic field (see Fig.\ref{plaquette}, 
\begin{equation}
\ket{m,n}  \overset{\square}{\longrightarrow}e^{i\left(A\left(m,n\right)-A\left(m,n+1\right)\right)} \ket{m,n}
\end{equation}

However if the atom starts the closed path at an adjacent site $\left(m+1\right)$ it will acquire the opposite phase because the Rabi frequency for the $\ket{g} \rightarrow \ket{e}$ is the complex conjugate of the Rabi frequency for the $\ket{e} \rightarrow\ket{g}$ transition. In other words, an atom going from an even to an odd row will pick up the opposite phase than an atom going from odd to even rows. This is the most important drawback of the scheme, as it leads to a staggered magnetic field. In order to remedy the problem, the authors of Ref. \cite{JZ03} proposed to break the symmetry between both transitions by tilting the lattice, which is very challenging from an experimental point of view. Recently, a way around this problem was proposed in Ref. \cite{DG09}. There, the authors propose the introduction of an optical superlattice which breaks the degeneracy between transitions in adjacent plaquettes. Because the degeneracy is broken, it is possible to drive, e.g. the transitions $\ket{m,n} \rightarrow \ket{m,n+1}$ and $\ket{m,n} \rightarrow \ket{m,n-1}$ with \textit{different} lasers, whose wave vectors can be chosen opposite to each other and thus rectify the staggered magnetic field. 

The scheme we propose to create a vector potential in different gauges requires a modification if it is implemented with an optical superlattice. It is necessary that the lasers that drive the different transitions, which were originally degenerate, be rotated simultaneously and by the same angle.

\section{Artificial gauge transformations}

In this section we discuss how the geometry of the Raman lasers applied in order to induce the artificial gauge field can be varied in order to transform the vector potential into another gauge. 

The Hamiltonian en Eq. \ref{hubbard} is invariant under a global transformation where all the creation and annihilation operators are replaced according to $a^{\dagger}_{m,n} \rightarrow e^{-i\beta} a^{\dagger}_{m,n}$ and $a_{m,n} \rightarrow e^{i\beta} a_{m,n}$. This leads to global particle number conservation. However, if the phase $\beta$ is allowed to vary from one lattice site to another, the Hamiltonian is no longer invariant. 

Nevertheless, it is possible to make this local transformation a symmetry of system by changing the configuration of the Raman lasers that generate the vector potential. Let $\bi{A}_1$ be the vector potential induced by a given configuration in which the Raman-lasers' wave-vectors form a given angle $\theta_1$ and frequency such that $\alpha=\alpha_1$. Then, it is straightforward to see that there always exists another configuration that gives rise to another vector potential $\bi{A}_2$ given by $\theta_2$ and $\alpha_2$ such that $\alpha_1\cos \theta_1=\alpha_2\cos \theta_2$, i.e. such that the magnetic flux is the same. Hence, by tuning $\alpha$ and $\theta$ accordingly it is possible to transform the vector potential into a whole family of gauges which differ by a linear function of $n$. Let us stress that these gauge transformations are forced upon the system, they \textit{do not} possess a dynamical origin.

The spectrum of the Hamiltonian in Eq \ref{hubbard2} is invariant under a broader set of gauge transformations \cite{LiebLoss}. Indeed, one can see that the Hamiltonian remains unchanged if the hopping phases, i.e. the artificial vector potential, is changed according to 

\begin{equation}
\eqalign{&A_x(m,n) \rightarrow A_x(m,n) - (\Lambda(m+1,n) - \Lambda(m,n))\cr
 &A_y(m,n)\rightarrow A_y(m,n) - (\Lambda(m,n+1) - \Lambda(m,n))\, ,}
\label{eq2} 
\end{equation}
as long as one redefines the creation and destruction operators following
\begin{equation}
\eqalign{ &a_{m,n} \rightarrow e^{i\Lambda(m,n)}a_{m,n} \cr
 &a^{\dagger}_{m,n} \rightarrow e^{-i\Lambda(m,n)}a^{\dagger}_{m,n} \, .}
\label{eq3} 
\end{equation}

A pure gauge configuration is one that does not contribute to the magnetic flux going through the system. Due to the fact that the spectrum of the Hamiltonian can only depend on the gauge field through the flux, a gauge invariant configuration can also be defined as one that has no physical effect. It follows that all pure gauge configurations can be cast into the form
\begin{equation}
A_{PG}=\left(\Lambda\left( m+1,n\right) -\Lambda\left( m,n\right) ,\Lambda\left( m,n+1\right) -\Lambda\left( m,n\right) \right) \,,
\end{equation}
as it is easy to check that $\oint_C A_{PG}=0$, where $C$ is an arbitrary closed path on the lattice.

It is particularly simple to transform the zero magnetic field $A_y=0$ configuration into a pure gauge. Indeed, for any $\alpha \neq 0 $ if one sets $\theta=\pi /2 $ the vector potential becomes $A'_y=\alpha n /2$, which is a pure gauge configuration. A test of the invariance of the system under artificial gauge transformations would be to perform the appropriate measurements on the system and verify that the same results are obtained for $A_y$ and $A'_y$.

\section{Insulator-Superfluid quantum phase transition}

In this section we will discuss the phase diagram of the Bose-Hubbard model, which is gauge invariant. This will allow us to give an experimental protocol to measure these quantities and hence test the gauge invariance of the actual system in a lab. 

A fundamental difference between systems coupled to real gauge fields and systems coupled to synthetic gauge fields is that in the first system only gauge invariant quantities are physical whereas in the second system this is not the necessarily case. Indeed, in Ref. \cite{YCPPPS09} the authors report having experimentally distinguished between two \textit{pure gauge} configurations. This is only possible if the measurements they performed were not gauge-invariant. While a gauge-dependent quantity in the ultracold gas system may be physically meaningful, it cannot have a counterpart in a system with charged particles coupled to a real gauge field.

It is well known that the BH Hamiltonian shows a phase transition in its ground state which depends on the ratio $J/U$. For $J/U\rightarrow0$ and commensurate filling the ground state is a product state with a definite occupation at each site,
\begin{equation}\label{mott_gs}
\ket{MI} = \bigotimes_{i=1}^{N_L} \ket{\bar{n}}_i\,
\end{equation} 
where $\bar{n}=N/N_L$ is the ratio between the total number of atoms,$N$, and the total number of lattice sites is $N_L$. The subscript $i$ labels the lattice site. In the other limit, $J/U\rightarrow\infty$ the ground state a product of $N$ single-particle Bloch functions with momentum corresponding to the minimum of the first energy band. In the limit where the system is very large both in number of atoms and lattice sites, keeping $\bar{n}$ fixed, the ground state can be shown to be a product over coherent single-site occupation states,
\begin{equation}\label{sf_gs}
\ket{SF}=\bigotimes_{m,n=1}^{L} $exp$\left(\sqrt{\bar{n}}a^{\dagger}_{m,n}\right)\ket{0}\,.
\end{equation}
where $N$ is the number of atoms. It is possible to distinguish between these two ground states by the presence or not of long range order or equivalently by computing the condensate fraction

\begin{equation}
\rho_0\equiv \left< a^{\dagger}_{k=0}a_{k=0} \right> = \frac{1}{N_L^2}\sum_{m,m',n,n'} \left\langle a^{\dagger}_{m,n} a_{m',n'} \right\rangle \, .
\end{equation}
which counts the number of atoms in which are in the zero momentum single particle state. As $J/U$ decreases the condensate is depleted until the many-particle reaches the Mott-Insulator state for $J/U=0$. Hence the phase transition between the two states is continuous.

A large condensate fraction is a signature of the superfluid phase, as it possesses long range order. Indeed, in the thermodynamic limit the only way $\rho_0$ can be large is by manifesting long-range non-vanishing correlations. Conversely, the Mott-state does not present long range order. Off-diagonal correlations decay exponentially fast, and $\rho_0$ is vanishingly small in the limit of an infinite lattice.

The location of the critical point between the two phases cannot be changed by a gauge transformation, as we shall discuss later on. If the order parameter is zero, i.e. the system is an insulator, it is zero independently of the choice of gauge. Indeed, the correlators in any two gauges will at most differ phase exp$\left(i\left(\Lambda\left({\bf{R}}\right)-\Lambda\left({\bf{R}}'\right)\right)\right)$.

\section{Determination of the ground state in mean field theory}

We now turn to the study of the superfluid-insulator phase diagram of the BH model in a magnetic field. This problem has been approached using strong-coupling perturbation theory in \cite{NFM99}, although we proceed with variational techniques.

An arbitrary state of the Hilbert space is given by
\begin{equation}
\ket{\Psi}=\sum_{n_1,n_2,..,n_N} C^{n_1,n_2,..,n_N} \ket{n_1,n_2,..,n_N}
\end{equation}
where $n_i$ is the atom number at site $i$ and $L^2$ is the total number of sites on the lattice. If we allow at most $d$ atoms to occupy each site, then the dimension of the many-body Hilbert space, which is the tensor product over all the single-site Hilbert spaces will have dimension $d^{L^2}$. The exponential dependence of the many-body Hilbert space dimension makes it impossible to attempt an exact diagonalization or a variational approach on the complete Hilbert space already for a $4 \times 4$ lattice with $d=3$. Consequently, an approximate scheme is needed where most of the states of the Hilbert space are ignored.

The fact that in both limits, $J\gg U$ and $J\ll U$, the ground state can be expressed as a product over single-site states allows for the quantum phase transition to be probed by a mean-field ansatz, also known as Gutzwiller's ansatz, which neglects quantum correlations between neighboring sites, 

\begin{equation}
\ket{\Psi_{MF}}=\bigotimes_{m,n}\sum_{N} f^N_{m,n} \ket{N}_{m,n}\, . 
\label{psimeanfield}
\end{equation}
This approximation is completely inadequate in one dimension, whereas it is exact in infinite dimensions. It was used in \cite{SKR93} to determine the phase diagram of the BH model in $2+1$ dimensions.

The mean field energy $E_{MF}$ is a sum of single site energies,

\begin{equation}
E_{MF}\left(\left\lbrace f^N_{m,n} \right\rbrace \right)=\sum_{m,n} E_{m,n}\left(\left\lbrace f^N_{m\pm 1,n},f^N_{m,n\pm 1}\right\rbrace \right)\, ,
\end{equation}
where $E_{m,n}$ depends only on the single-site states neighboring with site $(m,n)$. The authors of \cite{OT07} approximate the ground state of the BH model with a magnetic field by sequentially minimizing the local quantities $E_{m,n}$ with respect to the parameters $\left\lbrace f^N_{m\pm 1,n},f^N_{m,n\pm 1}\right\rbrace$ using a self-consistent method. However, in stead of using the general mean-field ansatz of Eq. \ref{psimeanfield}, they impose the $q$-periodicity of the Hamiltonian when the flux per plaquette, in our notation $\alpha\cos\theta=p/q$ is a rational number. Imposing this periodicity and translational invariance in the other direction on the trial state reduces the problem of minimizing $E_{MF}$ on an infinite lattice to that on a $q \times 1$ lattice.

In Ref. \cite{OT07} the optimization of the ansatz is carried out by defining a local Hamiltonian $H_{m,n}$ which only depends on the local states in the neighboring sites,  $\left\lbrace f^N_{m\pm 1,n},f^N_{m,n\pm 1}\right\rbrace$. Then $H_{m,n}$ is diagonalized in the occupation basis and the lowest lying eigenvector becomes the updated state at site $\left(m,n\right)$.

However, if the above method is applied to an unconstrained mean-field ansatz, where the symmetries are not imposed by hand, it often converges to local minima and there is no guarantee that the energy is being minimized at each step. To bypass this problem we optimize $\ket{\Psi_{MF}}$ by euclidean evolution.

\begin{figure}
\begin{center}
\scalebox{1}{\includegraphics{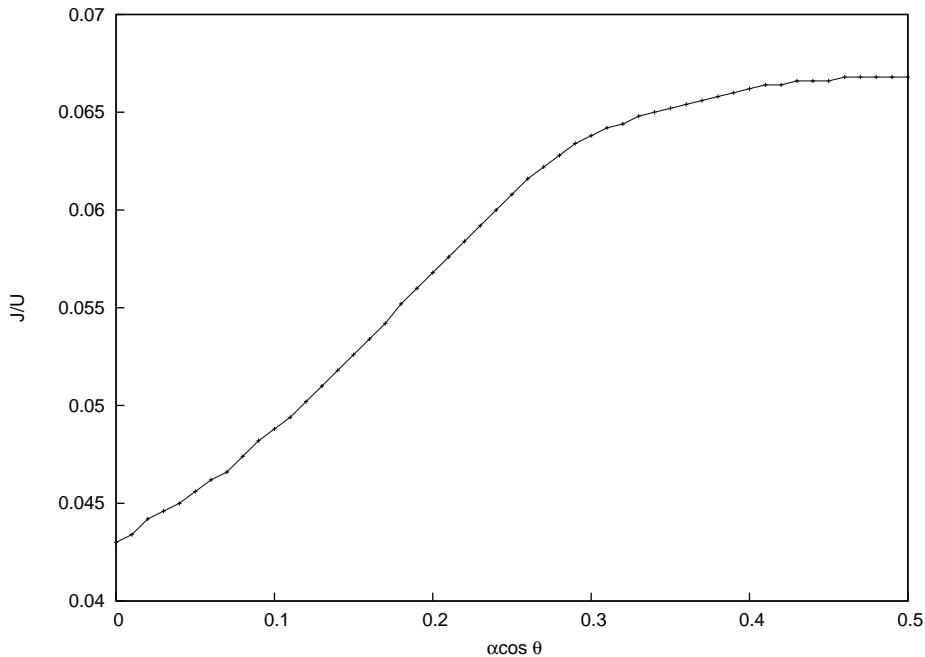}}
\end{center}
\caption{Critical value of $J/U$ for which the ground state of the BH Hamiltonian, Eq. \ref{hubbard2}, undergoes an insulator-to-superfluid transition as a function of the magnetic flux
$\Phi=\alpha \cos\theta$. The phase diagram was determined using the
mean field approximation for a $10\times 10$ square lattice with open boundary
conditions for $\mu_{m,n}=0.5U$ maximum one-site occupation $d=2$. As the angle of external lasers is
changed, a transfer from the transverse component to longitudinal component of the effective gauge field takes place and modifies the location of the critical value of $J/U$. For $\alpha \cos \theta
\rightarrow 0$, the hopping phase is $A_y(m,n)=\alpha n/2$,
and the Mott-insulator-to-superfluid transition point appears at the known value $(J/U)_c=0.043$.
Although site-dependent phases are present, this transition point remains
unchanged  because $A_y(m,n)$  is completely gauged-away, as it only depends
on $n$ and hence the derivative in the $m$ direction are zero. Only the interval $\Phi\in\left[0,0.5\right]$ is considered because the Hamiltonian in Eq. \ref{hubbard2} is invariant under the replacement $\Phi\rightarrow 1-\Phi$ and is periodic in $\Phi$ with unit period. }
\label{phasediagram}
\end{figure}

Fig. \ref{phasediagram} depicts the phase diagram of the model determined by optimizing the mean field ansatz with euclidean evolution. A particularly simple confirmation of the gauge-invariance of the model would be to check that the location of the Mott-to-Superfluid transition for a vector potential $A_y(m,n)=\alpha n/2$ is located at the same value of $J/U$ as for the case $A_y(m,n)=0$, as they are gauge equivalent because the former is a pure gauge configuration. In principle this can be done for any value of $\theta$, but the case $\theta=0$ is particularly simple.

\section{Determination of the ground state using a Tensor Network Ansatz}
The Hamiltonian under study, Eq. \ref{hubbard2}, presents the so-called phase problem which hinders progress 
through Monte Carlo calculations. Hence, in order to obtain results beyond mean-field we resort to a tensor 
network approach, which has proven to be very successful in describing one dimensional systems \cite{white1992}. 
Recently, a great deal of effort has been made towards developing a tensor network ansatz for two dimensional 
systems. In this section we employ one of these ans\"atze, namely the PEPS 
ansatz \cite{VC2004,SMD98,NO98}, to determine the phase diagram allowing for some degree of quantum correlation between 
different sites of the lattice. The ansatz replaces the quantities $f_n^i$ in each lattice site with tensors 
$A^{n_i}_{\alpha\beta\gamma\delta}$ where the Greek indices can take up to $\chi$ values. The trial state now 
reads

\begin{equation}
\ket{\Psi_{PEPS}}=\mathcal{F}\left(A^{n_{1,1}},\cdots, A^{n_{1,L}}, \cdots, A^{n_{L,L}}  \right) 
\ket{n_{1,1}\cdots n_{1,L}\cdots n_{L,L} } \, , 
\end{equation}
where the operator $\mathcal{F}(\cdot)$ stands for the contraction of the whole tensor network according to the 
connectivity of the lattice (in our case a square one) and the indices $n_{i,j}$ are summed over.

In order to compute local quantities all the indices of the tensors must be contracted. This can be done, 
approximately, in an efficient way. This means that the amount of resources needed to carry out the 
contraction scales polynomially in $\chi$. 

We will follow the algorithm described in \cite{MVC07} which is based on two different ways of finding the 
ground state of a given Hamiltionian. The first approach consists in iteratively minimizing the expected value 
of the Hamiltonian (the so-called direct minimization). The second one is called euclidean evolution and consists
in evolving a given PEPS state in imaginary time by minimizing the distance between the evolved state and a
new PEPS state. The computational effort is the same in each case and grows as $\chi^{12}$.

The implementation of the algorithm to the present work has been done by combining both methods. The direct
minimization is used for initializing the state and then the approximate ground state is reached by an 
evolution in imaginary time. Furthermore, we take the $\chi=1$ PEPS (mean-field solution) as the initial state 
for the $\chi=2$ PEPS in order to accelerate the convergence. Every converged PEPS is obtained with imaginary
time step $dt=10^{-3}$ and the Suzuki-Trotter expansion of second order. 

\begin{figure}
  \begin{minipage}[b]{5 cm}
 \begin{center}
   \includegraphics{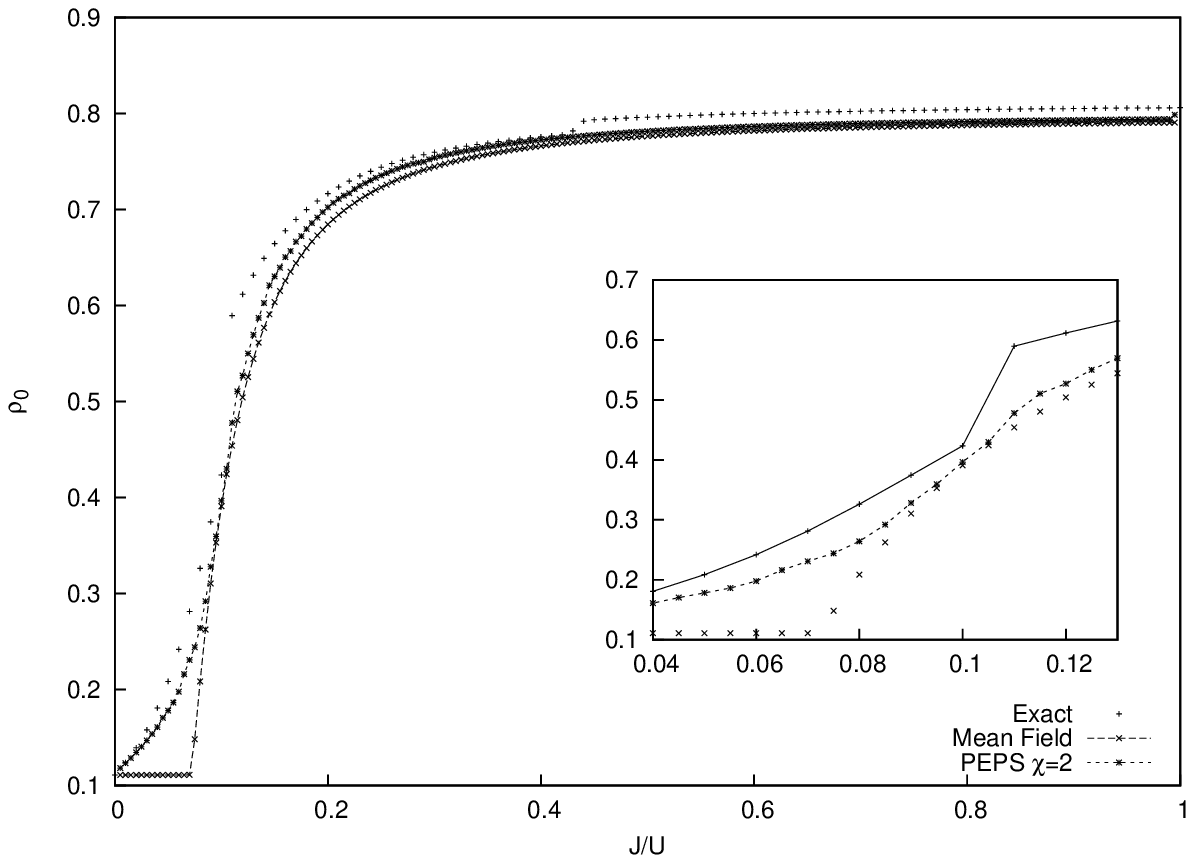}  
    \includegraphics{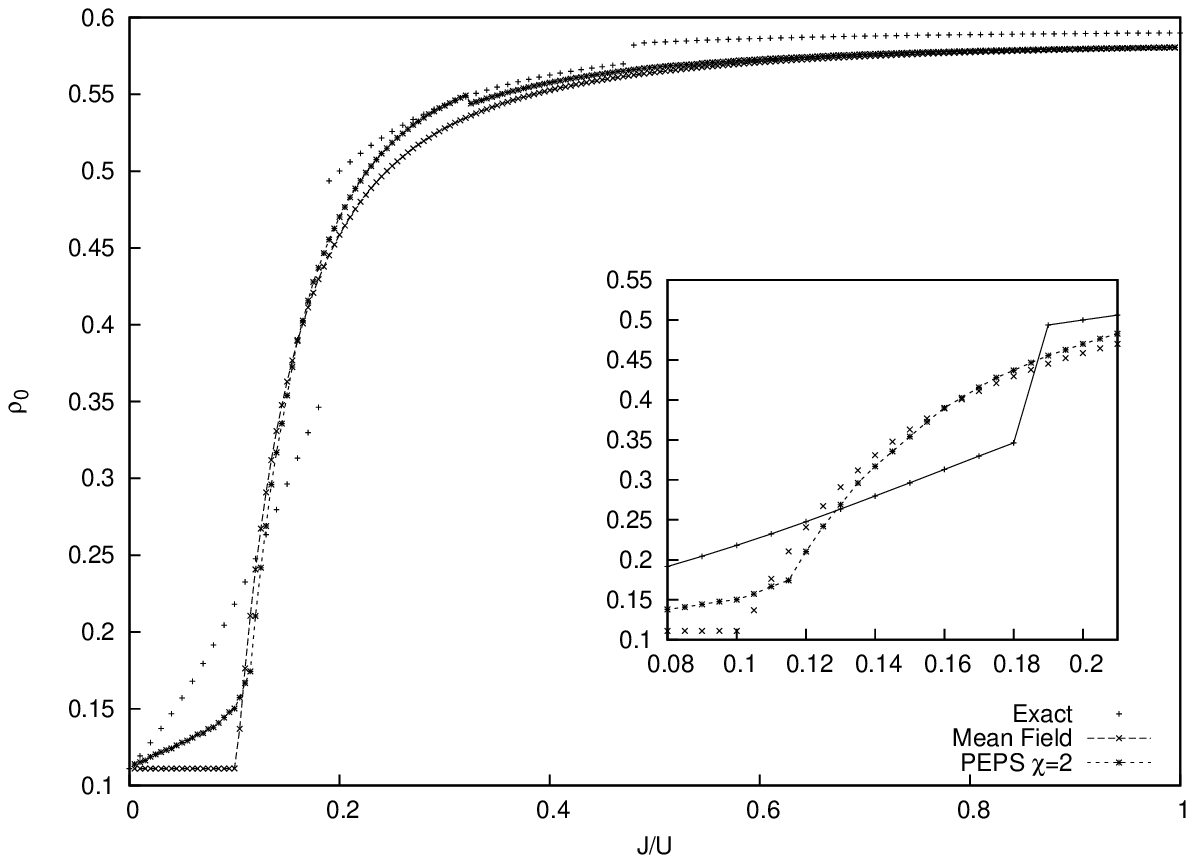}  
\end{center}
  \end{minipage}
 \caption{Condensate fraction $\rho_0$ as a function of $J/U$ for $\mu_{m,n}=0.5U$ and for two different values of magnetic flux $\Phi=\alpha\cos\theta=0$ (top) and $\Phi=0.5$ (bottom) computed by exact diagonalization and by PEPS ans\"atze with bond dimension $\chi=1, 2$ on a $3 \times 3$ lattice with maximum local occupation $d=2$. In all cases, $\rho_0$ tends to $1/N$ for small enough values of $J/U$ and approaches a macroscopic value for $J/U\rightarrow 1$, which is a sign of long range order. In order to compute the critical point we determine the value of $J/U$ where the variation of $\rho_0$ reaches its maximum. As can be seen with the naked eye, the critical point for $\alpha=0$ is located at a smaller $J/U$ than that for $\alpha=0.5$. The discontinuities in the exact curves are due to changes in total particle number of the ground state and are smoothed out in the thermodynamic limit.}
    \label{order_parameter}
\end{figure}

We compute the condensate fraction for different values of $\alpha$ for a $3\times3$ lattice of 3-valued sites, 
\textit{i.e.} sites whose occupation may be between 0 and 2. We choose two different values of alpha, 
0 and 1/2, in order to compare the two curves. Results are shown in Fig. \ref{order_parameter}.

For an infinite system the insulator-to-superfluid quantum phase transition is located at the value of $J/U$ for which the condensate fraction computed in the ground state becomes larger than $1/N$. However, for finite systems there is no exact phase transition. In order to estimate the position of the critical value of $J/U$ as accurately as possible for the infinite system from studying a finite system we compute the derivative of the condensate fraction with $J/U$ and determine the point where it is greatest. As the size of the lattice increases this criterion yields the correct result for the location of the critical point.

\section{Results}

The two phases of the infinite BH model in Eq. \ref{hubbard2} can be distinguished using as order parameter the condensate fraction $\rho_0$. In the superfluid phase corresponding to large $J/U$, the condensate fraction is non-zero because the ground state presents long range order. Conversely, in the Mott-insulator phase there is no such order and $\rho_0=0$. In the case of a finite size system, the transition does not appear at the point where $\rho_0$ is no longer vanishing but, rather, as a sharp transition from a small to a large value of $\rho_0$. The point of the phase transition is then determined as the point where the derivative of the order parameter reaches a maximum.  

In Fig.\ref{phasediagram}, we plot the phase diagram in the $J/U$ \textit{vs} $\Phi=\alpha\cos\theta$ plane,
that is, we analyze the value of the critical coupling $J/U$ as a function of the flux per plaquette. The system we have studied corresponds to a $10\times 10$ lattice with possible local occupations $d=0,1,2$. The result shown in this plot has been obtained by optimizing a mean field ansatz, which is a PEPS ansatz with $\chi=1$. Let us note
that, for $\Phi=0$, we recover the well known mean-field value for the quantum phase transition $J/U=0.043$. 
Then, as the magnetic field increases the critical value of $J/U$ raises monotonically until roughly $J/U=0.07$. 
This result produces a clear experimental prediction. The action of external laser fields as described in 
the previous chapters should shift the transition point to larger values than those measured in the absence
of effective magnetic fields.

In order to analyze whether this result is an artifact of the mean field approximation, we have studied
a smaller $3\times 3$ lattice using $\chi=1$ (mean field) and $\chi=2$ tensor networks as well as an exact diagonalization. Again, the possible occupation per site is $d=0,1,2$ and the chemical potential is maintained equal to $\mu=0.5 U$, which ensures that the Mott state has $\bar{n}=1$. The main result we obtain is that a $\chi=2$ PEPS does not differ significantly from that obtained with a $\chi=1$ ansatz, and both of them are very close to the exact result. We conclude that the phase diagram of the BH model obtained by variational means with a product state ansatz does not change significantly when a small amount of entanglement is included. The consistency between mean field and $\chi=2$ PEPS results holds for 
all values of the flux, though the departure of $\chi=2$ results from the $\chi=1$ ones is larger when
$\Phi=\alpha \cos \theta$ increases. 

\section{Conclusion}

Artificial non-dynamical gauge fields can be simulated in optical lattices using
external laser fields. The model describing such a system corresponds to a Bose-Hubbard
Hamiltonian with complex hopping phases.  
 
In this paper, we have presented a scheme that allows for the transformation of 
these artificial gauge fields. In particular, 
we have found the way to experimentally transform the vector potential into different gauges. 
We stress that these gauge transformations are unrelated to the gauge symmetry of the underlying electromagnetic interactions. 

We have also analyzed the phase transition present in this modified BH Hamiltonian using a PEPS ansatz.
We have determined the position of the Mott-insulator-to-superfluid critical point as a function of the magnetic flux going through the system. Our main result is that the critical value of $J/U$ of the insulator-to-superfluid phase transition grows smoothly with the induced magnetic flux. This result is already 
present in the mean field approximation and remains stable when PEPS are used. The modifications
introduced by the PEPS ansatz are more significant for large values of the induced external flux.  

\ack
The authors thank M. Lewenstein for useful discussions and comments. Financial support from QAP (EU), MICINN (Spain), Grup consolidat (Generalitat de Catalunya), and QOIT Consolider-Ingenio 2010 is acknowledged. O.B. was supported by FPI grant number BES-2008-004782.

\appendix
\section{}
\begin{figure}
\begin{center}
\begin{tabular}{lcr}
\includegraphics[scale=0.6]{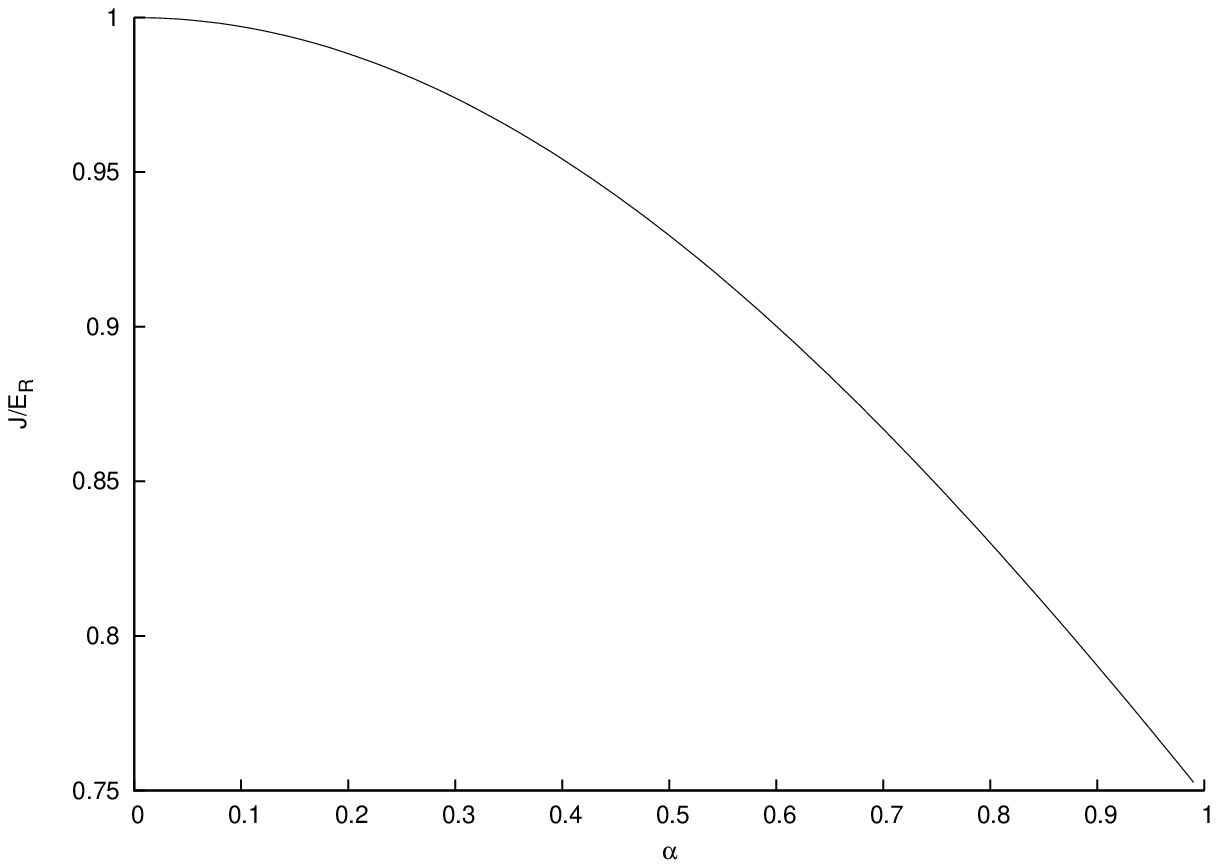} & \includegraphics[scale=0.6]{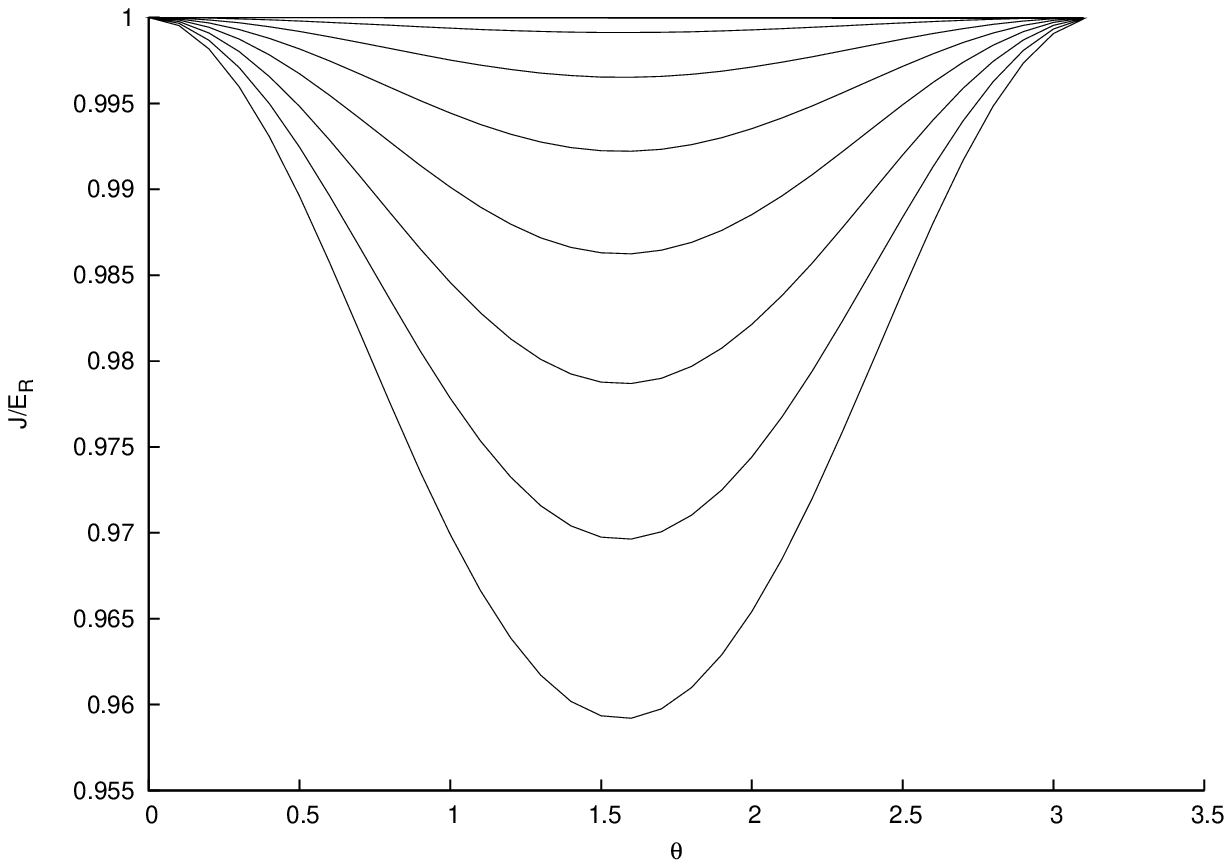} 
\end{tabular}
\caption{Running of $J_y(\alpha,\theta)$ for fixed $\theta$ (left) and for fixed $\alpha$. Clearly the variation is markedly larger in the former case than in the latter. It must be stressed that in both cases, the effective flux felt by the atoms on the lattice varies from its minimum to its maximum value.}
\label{Jalphatheta}
\end{center}
\end{figure}

In the calculation of the Raman-assisted tunnelling rate $J_R=Je^{iA_y\left(m,n\right)}$ we neglected to observe that the constant $J$ depends on $\alpha$ and $\theta$. In order to justify this assumption we numerically compute

\begin{equation}
\eqalign{
J = \frac{\Omega_0}{2} &\int  d x \vert \omega_x\left(x\right) \vert^2 \cos\left(2\alpha\cos\theta x\right)\times  \cr
 &\times\int  d y \omega_y^{*}\left(y\right)e^{2\pi i \alpha\sin\theta }\left(\bf{x}\right)\omega\left(y-\frac{\pi}{2}\right) \,.}
\end{equation}
As can be seen in Fig. \ref{Jalphatheta} the dependence on $J$ on $\alpha$ in units of the recoil energy $E_R$ is quite more significant then its dependence on $\theta$. This suggests that, in order to vary the magnetic flux going through the system, it would be far more efficient to vary $\theta$ in stead of $\alpha$. Indeed, to ensure that the system remains isotropous it is necessary to tune the value of $\Omega_0$ such that for every value of $\alpha$ we have $J_x=J$. By changing $\theta$ this compensation is not necessary because the maximum variation of $J$ as a function of $\theta$ is $4\%$, as opposed to $25\%$ when $\alpha$ is varied.

\end{document}